\documentclass[showpacs,preprintnumbers,amsmath,amssymb,      
twocolumn, tightenlines,prl
]{revtex4}

\usepackage[dvips]{graphicx}
\usepackage{bm}
\usepackage{amsmath}
\usepackage{dcolumn}

\begin{document}

\bibliographystyle{apsrev}

\title{Causality condition and reflection on event horizon}

\author{M.Yu.Kuchiev} \email[Email:]{kuchiev@newt.phys.unsw.edu.au}
\author{V.V.Flambaum} \email[Email:]{flambaum@newt.phys.unsw.edu.au}

    \affiliation{ School of Physics, University of New South Wales,
      Sydney 2052, Australia}
    
    \date{\today}

    \begin{abstract} 
 
      A new way to implement the causality condition on the event
      horizon of black holes is discovered. The metric of a black hole
      is shown to be a function of the complex-valued gravitational
      radius $r_{\mathrm{g}}\rightarrow r_\mathrm{g}+\,i0$. The
      relation between this modification of the metric and the
      causality condition is established using the analyticity of the
      S-matrix, which describes scattering of probing particles on a
      back hole. The found property of the metric has strong
      manifestations in scattering and related phenomena. One of them
      is the unexpected effect of reflection of incoming particles on
      the event horizon, which strongly reduces the absorption cross
      section.

    \end{abstract}
    

    \maketitle
    
    It is shown that quantum behavior of a probing particle on the
    event horizon of black holes should be described by the metric,
    which necessarily includes a small imaginary part. The causality
    condition and the analyticity of the scattering matrix provide a
    foundation for this result.
    
    Consider scattering of probing particles on a black hole. The wave
    equation that describes this process is presumed to be well known.
    For example, scattering of scalar particles is described by the
    Klein-Gordon equation. However, in order to formulate the problem
    properly, it is necessary to establish the boundary conditions on
    the horizon. At the first glance this is a straightforward task
    because on the horizon the semiclassical approximation is valid.
    Relying on the classical result, which states that incoming
    particles cross the horizon smoothly, it seemed natural to assume
    that the incoming wave is completely absorbed by the horizon,
    producing no reflected wave in the process. This boundary
    condition was introduced in the pioneering works on the scattering
    problem
    \cite{press_teukolsky_74,starobinsky_73,%
      starobinsky_churilov_73,unruh_76,sanchez_1977}, being used later
    on in a large number of (apparently all) further developments on
    scattering, see the books and reviews
    \cite{frolov_novikov_98,chandrasekhar_93,fullerman_handler_matzner_88,%
      sanchez_97} and references therein. It was also used in related
    problems, e.g. the radiation of gravitational waves by perturbed
    black holes, see the review \cite{rezolla_02} and reference
    therein.
    
    Recent Refs.\cite{kuchiev_1,kuchiev_2,kuchiev_3} pointed out
    though that thus formulated boundary condition is inconsistent,
    claiming that some part of the incoming wave is necessarily
    reflected on the horizon back into the outside world, i.e. the
    horizon is capable of reflecting particles.  This effect is
    referred to below as the reflection on the horizon (RH).  The RH
    was derived for eternal black holes using several different
    approaches in
    \cite{kuchiev_1,kuchiev_2,kuchiev_3,kuchiev_flambaum_4},
    Ref.\cite{kuchiev_flambaum_5} gives their brief summary, and for
    the collapsing case \cite{kuchiev_6}.
    
    A qualitative contradiction that exists between the RH and the
    conventional point of view, which assumes that the horizon is a
    perfect absorber, makes it necessary to study the problem in more
    detail. The present work suggests a new approach to the problem
    based on a new way of implementing the causality condition on the
    horizon. It shows that the metric, which governs the propagation
    of a probing particle on the horizon, should be considered as a
    complex-valued function of the modified gravitational radius
    $r'_\mathrm{g}$,
      \begin{eqnarray}
        \label{rg}
        r_{\mathrm{g}} \rightarrow  r'_\mathrm{g} =
        r_\mathrm{g}+\,i\gamma~,
        \quad \gamma \rightarrow +0~,
    \end{eqnarray}
    which has a small positive imaginary part. Here $r_\mathrm{g}=2
    GM$, $M$ is the mass of the black hole. The relativistic units
    $\hbar=c=1$ are used, if not stated otherwise. This modification
    of the gravitational radius proves very important for quantum
    phenomena, leading in particular to the RH, but is irrelevant in
    the classical approximation.
    
    Let us present first an example, which demonstrates the meaning
    and importance of Eq.(\ref{rg}). Consider for simplicity the
    Schwarzschild geometry described by the metric $ ds^2 = -\left(1
      -r'_\mathrm{g}/r \right)dt^2 + dr^2/ (1-r_\mathrm{g}' /r) + r^2
    d\Omega^2$ modified according to Eq.(\ref{rg}).  Consider a
    massless scalar particle with zero orbital momentum $l=0$ in the
    vicinity of the black hole. The corresponding Klein-Gordon
    equation for the radial wave function $\phi= \phi(r)$, which
    describes the state with the energy $\varepsilon$ reads
      \begin{eqnarray}
        \label{phi''}
        \phi''+\left( \frac{1}{r}+\frac{1}{r-r'_\mathrm{g}}\right)\phi'
        +\frac{\varepsilon^2 r^2}{ (r-r'_\mathrm{g})^2}\,\phi=0~.
    \end{eqnarray}
    It is convenient to introduce also the scaled wave function
    $\chi=[r(r-r'_\mathrm{g})]^{1/2}\,\phi$, which satisfies the wave
    equation
      \begin{eqnarray}
        \label{chi}
      \chi''+\frac{1}{(r-r_\mathrm{g}')^2}
      \left(\varepsilon^2r^2+\frac{r_\mathrm{g}'^2}{4r^2}
        \right)\chi =0~,
    \end{eqnarray}
    that looks similar to conventional Schr\"odinger-type equations
    $\chi''+(\varepsilon^2-U)\chi=0$ with an effective,
    energy-dependent potential $U$, which is defined so as to comply
    with Eq.(\ref{chi}). The term $i\gamma$ that comes in
    Eqs.(\ref{phi''}),(\ref{chi}) from (\ref{rg}) should be treated as
    finite at the intermediate stages of the analysis, the limit
    $\gamma \rightarrow 0$ is to be taken only at the end of
    calculations. This term regularizes the coefficients in the wave
    equation, making the equation finite at $r=r_\mathrm{g}$.
    (Moreover, we will argue below that it is convenient to consider
    this equation for negative values of the gravitational radius
    $(r_\mathrm{g}'<0)$, when it is not singular on the horizon, and
    derive the results for the physical value of $r_\mathrm{g}$ using
    the analytical continuation.)
      
    Consequently, Eq.(\ref{chi}) can describe the propagation of the
    wave function through the horizon. The fact that the outside
    observer cannot witness the crossing of the horizon by a particle
    presents no obstacles for the penetration of the wave function
    into the inside region. The causality principle, which forbids
    this crossing, works only for real physical processes, whereas the
    wave function by itself describes virtual processes.  In order to
    extract amplitudes of physical processes one has to fulfil some
    transformation on the wave function, which must ensure that the
    causality is preserved (see below).
      
    To make this point more transparent, consider a simple example,
    the conventional propagator, which describes the free scalar
    particle $D({\bf p},\varepsilon)=\left(\varepsilon^2-\epsilon_{\bf
        p}^2+i\gamma\right)^{-1}$, where $\epsilon_{\bf p}$ is the
    energy of the particle with the momentum ${\bf p}$. Transforming
    the propagator into the coordinate representation, one finds that
    it remains nonzero for positive intervals, $D({\bf r},t)\ne 0$ for
    ${\bf r}^2-t^2>0$. Consequently, the perturbation theory based on
    this propagator results in the wave function, which penetrates
    into those regions of space-time that appear forbidden by the
    causality condition. This fact, however, does not come into
    contradiction with the causality for physical processes due to the
    correctly chosen imaginary part $i\gamma$ in the propagator
    $D({\bf p},\varepsilon)$, see e.g.  the book
    \cite{bogolubov_shirkov_80}. Similarly, the fact that
    Eq.(\ref{chi}) allows the wave to propagate into the inside region
    cannot be discarded on the grounds of the causality principle.
    Generically, the wave function is allowed to propagate into any
    region, where the wave equation is well-defined. We will revisit
    this point later on, when discussing physical origins of
    Eq.(\ref{rg}).
      
    First, let us finish our discussion of the consequences of
    Eq.(\ref{rg}). Take the asymptotic solutions of Eq.(\ref{phi''})
    on the horizon $\phi=\exp[\mp i\,\varepsilon
    r_\mathrm{g}\ln(r/r'_\mathrm{g}-1)]$, $r-r_\mathrm{g}\ll r_g$,
    where the terms with minus and plus signs describe the incoming
    and outgoing waves respectively.  In the scattering problem it is
    convenient to start from the incoming wave in the outside region
      \begin{eqnarray}
        \label{phiIN}
        \phi_\mathrm{in}=\exp[-i\varepsilon r_\mathrm{g}'
        \ln(r/r'_\mathrm{g}-1)\,]~.
    \end{eqnarray}
    The finite term $i\gamma$, which is present in Eq.(\ref{phiIN})
    allows this wave to penetrate into the inside region, where it
    reads $\phi_\mathrm{in}=\rho \exp[- i\varepsilon
    r_\mathrm{g}\ln(1-r/r'_\mathrm{g})]$ for $r<r_\mathrm{g}$, being
    suppressed by a factor $\rho= \exp(-\pi\varepsilon
    r_\mathrm{g})<1$.  This fact prompts one to formulate the boundary
    condition at the origin.  Choosing the regular solution, which is
    conventional for the partial wave expansion, one concludes that in
    the inside region there necessarily exists also the outgoing wave,
    which should have the same amplitude as the incoming wave.  On the
    horizon of the inside region this outgoing wave reads
    $\phi_\mathrm{out}=\rho \exp[ \,i\varepsilon r_\mathrm{g}
    \ln(1-r/r'_\mathrm{g})+i\alpha]$.  Here $\alpha$ is the phase,
    which appears due to the propagation in between the horizon and
    the origin; its explicit value is irrelevant for the present
    study. Crossing the horizon (with the help of the same finite term
    $i\gamma$) this wave enters the outside region
      \begin{eqnarray}
        \label{out}
        &&\phi_\mathrm{out}=\mathcal{R} \exp[\,i\varepsilon r_\mathrm{g}'
        \ln(r/r'_\mathrm{g}-1)\,]~,
        \\ \label{R}
        &&|\mathcal{R}| = \rho^2
        =\exp\left(-\varepsilon/2T_\mathrm{H} \right)~.
   \end{eqnarray}
   Here $ T_\mathrm{H}= \hbar c/(4\pi r_\mathrm{g} )$ is the Hawking
   temperature (conventional units). Combining Eqs.(\ref{phiIN}) and
   (\ref{out}) one finds that the proper wave function, which
   describes the incoming particle in the outside region, reads
      \begin{eqnarray}
        \label{proper}
        \phi= \phi_\mathrm{in}+
      \mathcal{R}\,\phi_\mathrm{out}~. 
   \end{eqnarray}
   Here the first term represents the initial incoming wave, while the
   second one clearly describes the outgoing wave that emanates from
   the horizon. This means that particles are reflected on the
   horizon. This phenomenon is called the RH. The quantity $\mathcal
   R$, which describes the intensity of the RH, represents the
   reflection coefficient. Notably, it does not depend on $\gamma$
   introduced in Eq.(\ref{rg}), that is the necessary limit $\gamma
   \rightarrow 0$ is taken into account automatically.
   
   Eq.(\ref{R}), which was first derived in
   \cite{kuchiev_1,kuchiev_2,kuchiev_3}, shows that in the high energy
   region $\varepsilon > T_\mathrm{H}$ the reflection coefficient is
   small, making the RH insignificant. The classical result, which
   predicts that the RH is impossible, i.e.  $\mathcal{R}=0$, is
   recovered in a limit $\hbar \rightarrow 0$ that implies that
   $T_\mathrm{H}=0$. In contrast, in the infrared region $\varepsilon
   \ll T_\mathrm{H}$ the RH is very strong, $|\mathcal{R}|\simeq 1$,
   which makes the horizon a good infra-red mirror.
       
   In accord with the latter claim, the RH drastically suppresses the
   low-energy absorption cross-section
       \begin{equation}
        \label{sigma}
        \sigma_\mathrm{abs} = \eta\,
        \sigma_\mathrm{abs}^{(0)}~,\quad \eta=\pi
        \varepsilon r_\mathrm{g}\ll 1~. 
   \end{equation}
   Here $\sigma_\mathrm{abs}^{(0)}$ is the cross section, which
   neglects the RH, the factor $\eta$ accounts for the RH, reducing
   the absorption cross-section.  Eq.(\ref{sigma}) was first derived
   in Ref.  \cite{kuchiev_flambaum_4} for a particular case, the
   s-wave scattering of scalar particles. In the present work we
   verified by explicit calculations that Eq.(\ref{sigma}) is valid
   for the impact of scalar particles, photons and gravitons on
   Schwarzschild black holes for any allowed quantum numbers of these
   particles \footnote{Presumably Eq.(\ref{sigma}) remains applicable
     for fermions as well, though explicit calculations related to
     this fact has not yet been finished.}.
       
   After discussing the consequences of Eq.(\ref{rg}), let us now
   prove its validity. Our starting point is the causality condition.
   As is well known, in any quantum field theory the causality
   condition makes the scattering matrix an analytical function of
   energy \cite{bogolubov_shirkov_80}. One may argue that quantum
   gravity does not completely fall under the category of
   well-understood quantum field theories, questioning an
   applicability of the general theorem that relates the causality and
   analyticity. Fortunately, there exist independent methods for
   verification of analyticity.  Eq.(\ref{chi}) is similar to a
   conventional Schr\"odinger-type equation. One can rely therefore on
   the Landau argument discussed in \cite{landau_lifshits_77}, which
   states that the S-matrix is an analytical function of energy on the
   physical sheet, if the potential $U(r)$ satisfies condition $U(r)
   \rightarrow 0$, when $r\rightarrow \infty$ in the semiplane
   $\mathrm{Re}\, r >0$ of the complex plane $r$.  This is obviously
   the case for the potential $U(r)$ that governs Eq.(\ref{chi})
   \footnote{The effective nature of this energy-dependent potential
     bears no negative consequences for the applicability of the
     Landau argument because in our case the necessary condition
     $U(r)\rightarrow 0$ is valid when $|r| \rightarrow \infty$ for
     arbitrary direction on the complex plane $r$}.
   
   Thus, the general theorem based on the causality condition and a
   more specific approach, which relies on the Landau argument, both
   show that the S-matrix is an analytical function of energy. For
   simplicity we will continue to restrict ourselves to the most
   transparent case of the impact of massless scalar particles with
   $l=0$ on Schwarzschild black holes. Since the S-matrix is
   analytical on the physical sheet ($\mathrm{Im}\, \varepsilon \ge
   0$) \footnote{The variable $\varepsilon$ in Eq.(\ref{chi}) plays
     the role of the effective momentum $k$ for the Shr\"odinger
     equation; in terms of the momentum the physical sheet is defined
     as $\mathrm{Im}\,k\ge 0$.}, one can extend its definition as far
   as the negative semiaxis $\varepsilon <0$.
       
   Note now that the S-matrix also depends on parameters that govern
   the potential.  In our case there is only one such parameter, the
   gravitational radius. Thus, the S-matrix is a function $S =
   S(\varepsilon, r_\mathrm{g}')$. Moreover, this function should
   depend on dimensionless quantities only, which makes it a function
   of only one available dimensionless variable $\varepsilon
   r_\mathrm{g}'$
       \begin{eqnarray}
         \label{S}
         S\equiv S(\varepsilon r_\mathrm{g}')~.
   \end{eqnarray}
   One can treat the gravitational radius $r_\mathrm{g}'$ in
   Eq.(\ref{S}) as a complex-valued parameter. Combining the
   analyticity of the S-matrix over the energy with Eq.(\ref{S}) one
   immediately deduces that for positive physical energies
   ($\varepsilon>0$) the S-matrix is an analytical function of
   $r_\mathrm{g}'$ in the upper semiplane
   ($\mathrm{Im}\,r_\mathrm{g}'\ge 0$) of the complex plane
   $r_\mathrm{g}'$. Therefore, starting from the physical region
   ($\varepsilon > 0,~r_\mathrm{g}'> 0$), the S-matrix can be defined
   in the upper semiplane of the gravitational radius, which extends
   up to the negative semiaxis ($r_\mathrm{g}'<0$).
   
   Let us now revert the argument, using the analyticity of the
   S-matrix as a tool for proper definition of the scattering problem.
   Consider the range of parameters ($\varepsilon >0, ~r_\mathrm{g}' <
   0$).  The condition $r_\mathrm{g}'<0$ makes the coefficients of the
   wave equation Eq.(\ref{chi}) regular real functions of the radius
   $r$ in the physically allowed interval $0<r<\infty$. This fact
   greatly simplifies the boundary conditions, which state that the
   wave function should be regular at the origin $r=0$ and describe
   the free-type propagation at infinity $r=\infty$.  Since the
   equation is regular at $r=-r_\mathrm{g}'=|r_\mathrm{g}'|$, one does
   not need to formulate explicitly the boundary conditions on the
   horizon.
   
   Thus, the chosen range of parameters ($\varepsilon >0,
   ~r_\mathrm{g}' <0$) makes scattering an ordinary problem formulated
   on the interval $0<r<\infty$.  Solving it and extracting the
   corresponding S-matrix one can use Eq.(\ref{S}) to re-interpret the
   result, proclaiming that the S-matrix found describes the case of a
   positive gravitational radius and negative energy ($\varepsilon <
   0, ~r_\mathrm{g}' > 0$). Finally, using the analyticity of the
   S-matrix over the energy, one can continue it from the negative
   semiaxis of energies into the physical region, where both the
   energy and the gravitational radius are positive ($\varepsilon>
   0,~r_\mathrm{g}'>0$).
       
   We see that scattering in the physical region ($
   \varepsilon>0,~r_\mathrm{g}'>0$) can be described in terms of the
   scattering problem formulated for non-physical, negative
   gravitational radiuses ($\varepsilon>0,~r_\mathrm{g}'<0$), where
   the problem is much simpler because the coefficients of the wave
   equation are regular. Developing this argument, one observes that
   the coefficients in Eq.(\ref{chi}) remain regular analytical
   functions of $r_\mathrm{g}'$ in all of the upper semiplane
   ($\mathrm{Im}\,r_\mathrm{g}' > 0 $) for all values of the argument
   $r$ in the physical region $0<r<\infty$. Therefore, one can rely on
   this equation when formulating the scattering problem for any value
   of the gravitational radius in this semiplane, where the boundary
   condition on the horizon remains redundant. In particular,
   $r_\mathrm{g}'$ can be chosen in a vicinity of its physical value,
   which is exactly what Eq.(\ref{rg}) is suggesting.  We see that the
   purpose of Eq.(\ref{rg}) is to ensure that the wave equation and
   the S-matrix are defined via the analytical continuation from the
   upper semiplane of the gravitational radius $\mathrm{Im}\,
   r_\mathrm{g}'>0$, which ensures that the S-matrix is an analytical
   function in this area, as it should be.
   
   Let us briefly summarize the arguments unfolded above.  The
   starting point was the causality condition, which makes the
   S-matrix an analytical function of energy (more direct arguments
   support this fact). The analyticity over energy brings in the
   analyticity of the S-matrix over the gravitational radius,
   prompting one to consider the wave equation for complex values of
   the gravitational radius, which makes the coefficients in the wave
   equation finite, regular functions. As a result, the explicit
   formulation of the boundary conditions on the horizon becomes
   redundant. Returning to the positive physical value of the
   gravitational radius one then recovers the physical S-matrix via
   its analytical continuation.  Thus, the analyticity of the S-matrix
   is used as a tool for derivation of the proper boundary conditions
   on the horizon in the physical region.
   
   Apparently, the suggested modification of the metric should have
   serious implications for {\it any} quantum phenomenon near a black
   hole, as shows the considered example of scattering of scalar
   particles (electromagnetic and gravitational waves were also
   mentioned above, but only briefly). We demonstrated that the
   boundary conditions on the event horizon, which follow from
   Eq.(\ref{rg}), lead to the surprising effect of reflection of
   incoming particles from the horizon. The chain of arguments, which
   lead from the causality condition via the analyticity of the
   S-matrix to the reflection, makes this derivation transparent and
   reliable. The reflection on the horizon reduces the absorption
   cross section, which may be of interest for description of
   evolution of primordial black holes at early stages of the Big
   Bang.
      
   This work was supported by the Australian Research Council.


\begin{thebibliography}{99}
       

   \bibitem{press_teukolsky_74}
    
    W.H. Press and S. A. Teukolsky, Astrophys. J. {\bf 141}, 443 (1974).


  \bibitem{starobinsky_73} 
    
    A. A. Starobinsky, Zh. Eksp. Teor. Fiz.  {\bf 64}, 48 (1973)
    [Sov.Phys.JETP {\bf 37}, 28 (1973)].


  \bibitem{starobinsky_churilov_73} 
    
    A. A. Starobinsky and C. M. Churilov, Zh. Eksp. Teor. Fiz.  {\bf 65},
    3 (1973) [Sov.Phys.JETP {\bf 38}, 1 (1974)].


  \bibitem{unruh_76}
    
    W. G. Unruh, Phys. Rev. D {\bf 14}, 3251 (1976).


  \bibitem{sanchez_1977}
    
    N. Sanchez, Phys. Rev. D {\bf 18}, 1030 (1977).


   \bibitem{frolov_novikov_98}
      
     V. P. Frolov and I. D. Novikov, {\it Black hole physics: basic
       concepts and new developments } (1998) Dordrecht; Boston: Kluwer.


  \bibitem{chandrasekhar_93}
      
    S. Chandrasekhar, {\it The Mathematical Theory of Black Holes}
    (1993) New York: Oxford University Press.


  \bibitem{fullerman_handler_matzner_88} 
    
    J. A. H. Futterman, F. A.  Handler, and R. A.  Matzner,  {\it
      Scattering from black holes} (1988) Cambridge; New York:
    Cambridge University Press.


  \bibitem{sanchez_97}

    N. Sanchez, {\it The black hole: scatterer absorber and emitter of
    particles} hep-th/9711068.


  \bibitem{rezolla_02}
    
    L. Rezzolla, {\it Gravitational waves from perturbed black holes
      and relativistic stars}, gr-qc/0302025. ``Published in Trieste
    2002, Astroparticle physics and cosmology'', pp. 255-316.
   

  \bibitem{kuchiev_1} 
    
    M. Yu. Kuchiev, {\it Reflection from black holes}, \\ gr-qc/0310008.


  \bibitem{kuchiev_2}

  M. Yu. Kuchiev, Europhys. Lett. {\bf 65}, 445 (2004).


  \bibitem{kuchiev_3}
    
    M. Yu. Kuchiev, Phys. Rev. D {\bf 69}, 124031 (2004).


  \bibitem{kuchiev_flambaum_4}
    
    M. Yu. Kuchiev and V. V. Flambaum, Phys. Rev. D {\bf 70}, 044022
    (2004).


  \bibitem{kuchiev_flambaum_5}
    
    M. Yu. Kuchiev and V. V. Flambaum, {\it Reflection on event
      horizon and escape of particles from confinement inside black
      holes}, gr-qc/0407077.


  \bibitem{kuchiev_6}
    
    M. Yu. Kuchiev, {\it Reflection on event horizon for collapsing
      black holes}, gr-qc/0411009.


  \bibitem{bogolubov_shirkov_80}
    
    N. N. Bogoliubov and D. V. Shirkov, {\it Introduction to the
      theory of quantized fields} (1980) New York: Wiley.


  \bibitem{landau_lifshits_77}
      
    L. D. Landau and E. M. Lifshits, {\it Quantum mechanics:
      non-relativistic theory} (1977) New York: Pergamon.


  \bibitem{thooft_01} 
    
    G. 't Hooft, {\it Horizons}, gr-qc/0401027. Lecture presented at
    the Erice School for Sub-Nuclear Physics (2003).

    
  \bibitem{thooft_96} G. 't Hooft, J. Mod. Phys. A {\bf 11}, 4623
    (1996).


%


%


%


  \bibitem{hawking_74}
      
    S. W. Hawking, Nature (London) {\bf 248}, 30 (1974).


  \bibitem{hawking_75}
      
    S. W. Hawking, Commun. Math. Phys.  {\bf 43}, 199 (1975).


\end{thebibliography}
   \end{document}